\documentclass[%
 reprint,
 amsmath,amssymb,
 aps,
]{revtex4-2}

\usepackage{graphicx}
\usepackage{float}
\usepackage[utf8]{inputenc}
\usepackage{textgreek}
\usepackage{dcolumn}
\usepackage{xcolor}
\usepackage{bm}
\usepackage{bbm}
\usepackage{subcaption}
\usepackage{hyperref}
\hypersetup{
    colorlinks=true,       
    linkcolor=blue,        
    citecolor=blue,      
    urlcolor=blue}
    
\begin{document}

\preprint{APS/123-QED}

\title{Gravitational Wave in an $A_4 \otimes Z_4$ Neutrino Model with Generalised CP}

\author{Pulakesh Borah}
 \email{Electronic adress: \textcolor{magenta}{pborah651@gmail.com}}

\author{Mrinal Kumar Das}
 \email{Electronic adress: \textcolor{magenta}{mkdas@tezu.ernet.in}}
\affiliation{
 Department of Physics, Tezpur University, Napaam, Assam, India-784028}

\begin{abstract}

We extend the Standard Model with an $A_4 \times Z_4$ flavor symmetry and generalized CP (GCP) symmetry, realized through two flavon triplets that spontaneously break $A_4$ and generate the observed lepton mixing pattern. A new $A_4$-singlet flavon couples to one of these triplets through a complex quartic interaction that breaks GCP and fixes the reactor mixing angle; this same coupling biases the domain walls formed when that triplet breaks $A_4$. A global fit to oscillation data reproduces all mixing angles and mass splittings, predicting a maximal Dirac CP phase of $270^\circ$ and an effective Majorana mass near $15\text{ meV}$, testable by upcoming neutrinoless double-beta-decay ($0\nu\beta\beta$) experiments. The same coupling generates a stochastic gravitational-wave background peaking in the pulsar-timing-array and LISA bands, with peak frequency and amplitude predicted to scale as $\sin\theta_{13}$ and $\sin^{-4}\theta_{13}$, a falsifiable link between neutrino oscillation and gravitational-wave (GW) observations.

\end{abstract}

\maketitle

\section{\label{sec:level1}Introduction}

Although the Standard Model (SM) of particle physics successfully describes most fundamental interactions, the discovery of nonzero neutrino masses and large lepton mixing angles demands physics beyond the SM. While the seesaw mechanism offers a compelling framework for explaining the extreme smallness of light neutrino masses through heavy new degrees of freedom, the origin of the observed lepton flavor mixing pattern remains unexplained within the minimal framework. This contrast is especially striking in the quark and lepton sectors; quark mixing follows a strict, nearly diagonal pattern, whereas leptonic mixing has a different structure with two large angles and one small, non-zero reactor angle.

A promising approach to addressing this flavor puzzle involves extending the SM with non-Abelian discrete flavor symmetries such as $S_4$ \cite{ma2006neutrino, altarelli2009revisiting, altarelli2010discrete}, $A_4$ \cite{altarelli2006tri, ma2001softly, das2020phenomenological, pathak2026matter, kumar2025neutrino}, $A_5$ \cite{everett2009icosahedral, ding2012golden} and $\Delta(27)$ \cite{ma2007near} etc, which can naturally dictate the observed lepton mixing patterns and mass relations. Among these, $A_4$ has long been a primary focus. As the smallest group containing both a $\mathbb{Z}_3$ and a $\mathbb{Z}_2$ subgroup, it naturally accommodates the residual symmetries of the charged-lepton and neutrino sectors, respectively, and predicts tri-bimaximal mixing (TBM) at leading order \cite{chen2026gravitational}. Although TBM is now excluded by the measured nonzero reactor angle $\theta_{13}$ \cite{esteban2024nufit,nufit61}, $A_4$ remains a compelling starting point once supplemented by a mechanism generating a realistic deviation from this limit, commonly through cross-couplings between the flavons responsible for spontaneously breaking $A_4$ \cite{pascoli2016role}.

A generic and largely unavoidable byproduct of this spontaneous breaking is the formation of cosmic domain walls, arising whenever the vacuum manifold of the flavon potential is disconnected \cite{Zeldovich:1974uw}. If left intact, these walls would dominate the energy density of the universe, contrary to cosmological observations; a small explicit or spontaneous breaking of the residual discrete symmetry is therefore required to bias the vacua and trigger wall annihilation before nucleosynthesis \cite{chauhan2024phenomenology,gelmini1989cosmology}. The resulting collapse sources a stochastic GW background whose spectral shape and amplitude are set by the wall tension and bias energy \cite{hiramatsu2010gravitational,saikawa2017review}. Recently, it was shown that in $ A_4$-based flavor models the cross-couplings between flavons of the model supply the bias necessary to annihilate the domain wall, tying the frequency and amplitude of the predicted GW signal directly to the flavor sector, with a benchmark spectrum intersecting the stochastic signal reported by pulsar timing arrays such as NANOGrav \cite{chen2026gravitational}. 

In this work, we construct a model based on $A_4$ flavor symmetry, extended by an additional $\mathbb{Z}_4$ symmetry and a GCP symmetry imposed on the full Lagrangian. The scalar sector includes two $A_4$ triplet flavons, one of which governs the charged-lepton sector and the other the neutrino sector, together with an $A_4$ singlet flavon $\xi''$ carrying nontrivial $\mathbb{Z}_4$ charge. This singlet flavon couples to the neutrino-sector triplet through a single complex quartic interaction. GCP invariance forces every coupling in the Lagrangian to be real, with one exception: the phase of this quartic coupling cannot be removed by any field redefinition, since the residual rephasing freedom of the singlet flavon is entirely fixed by its other appearances in the Lagrangian. A nonzero complex phase in this coupling therefore constitutes the unique, explicit source of CP violation in our model. This GCP-violating phase fixes the vacuum phase of the singlet flavon, which generates the reactor mixing angle through the perturbative diagonalization of the neutrino mass matrix that determines the full lepton mixing pattern. At the same time, the same coupling lifts the degeneracy of the three $Z_2$-preserving vacua of the neutrino-sector flavon, biasing the associated domain walls and triggering their annihilation into a stochastic GW background. Because both effects trace back to a single coupling, this construction yields a model-specific correlation between the reactor mixing angle and the resulting gravitational-wave spectrum, offering a novel observational probe that connects the neutrino sector directly to the gravitational-wave background through a common origin, a connection previously absent in frameworks where these phenomena were treated independently.

We perform a global fit of the neutrino sector to NuFit 6.1 oscillation data \cite{esteban2024nufit,nufit61} in the Altarelli-Feruglio basis, and separately derive the domain-wall tension self-consistently from the Bogomolny-Prasad-Sommerfield (BPS) bound on the flavon potential \cite{fu2026domain}, rather than treating it as an independent free parameter. This self-consistency requirement uniquely fixes the absolute flavon symmetry-breaking scale while leaving every neutrino-sector observable completely unaffected, since they depend entirely on dimensionless mass and mixing ratios. We show that the resulting GW spectrum can populate both the pulsar-timing-array \cite{antoniadis2023second} and LISA \cite{colpi2024lisa} frequency bands. We explicitly derive the predicted correlation $f_\text{peak}\propto\sin\theta_{13}$ and $\Omega_\text{peak}\propto\sin^{-4}\theta_{13}$, offering a concrete, falsifiable target for forthcoming reactor, long-baseline, and gravitational-wave experiments. 

Furthermore, beyond mixing angles and mass splittings, our framework makes a distinct, independently testable prediction for $0\nu\beta\beta$ \cite{jones2021physics,dolinski2019neutrinoless,bilenky2012neutrinoless,gomez2011sense}. Because the light neutrino masses are generated via the type-I seesaw mechanism within this setup, neutrinos acquire a Majorana nature, fixing the effective Majorana mass $|m_{ee}|$. As a lepton-number-violating process, the observation of $0\nu\beta\beta$ would confirm the Majorana character of neutrinos, while current and forthcoming experimental non-observations place stringent limits on the parameter space \cite{abe2025search}. The resulting prediction for $|m_{ee}|$, derived directly from our model's parameter space, offers an essential complementary cross-check alongside oscillation measurements.

This work is organized as follows. In Sec. II, we briefly review domain wall formation and the necessity of bias generation using a toy potential. Section III outlines all components of the model framework. The numerical analysis for the neutrino sector and neutrinoless double beta decay is presented in Sec. IV. Section V is dedicated to the study of gravitational waves arising from domain wall annihilation. Finally, we present our conclusions in Sec. VI.

\section{\label{sec:level2} Domain Wall and Necessity of Bias Generation}

When the Universe cools past the critical temperature of a symmetry-breaking phase transition, scalar fields settle into nonzero vacuum expectation values, and any discrete symmetry they carried gets spontaneously broken. Because causally disconnected patches of the Universe can't communicate across the horizon, different regions end up choosing different degenerate vacua. Wherever two such regions meet, a domain wall forms at the boundary.

The simplest way to see this is with a single real scalar field $\phi$ undergoing $Z_2$ breaking ($\phi \to -\phi$), governed by the familiar double-well potential of the form
\begin{equation}
    V(\phi)=\frac{\lambda}{4}(\phi^2 - v_\phi^2)^2
\end{equation}
where $\lambda$ is the dimensionless quartic coupling and $v_\phi$ is the symmetry-breaking scale. This potential possesses two degenerate minima at $\phi = \pm v_\phi$. If one region falls into $+v_\phi$ and its neighbor into $-v_\phi$, the field has to interpolate smoothly between them, and minimizing the energy gives the usual kink profile:
\begin{equation}
    \phi(x)=v_\phi \tanh(\sqrt{\frac{\lambda}{2}}v_\phi x)
\end{equation}

This configuration stores energy in the wall itself. Integrating the energy density across the profile gives the wall tension,
\begin{equation}
    \sigma=\int_{-\infty}^{\infty} (\frac{1}{2}(\partial_x \phi)^2 + V(\phi))\,dx = \frac{4}{3}\sqrt{\frac{\lambda}{2}} v_\phi^3 = f_\sigma v_\phi^3
\end{equation}

In the scaling regime, the network's energy density falls off as $\rho_{DW}=\frac{\sigma}{t}$ much more slowly than radiation ($a^{-4}$) or matter ($a^{-3}$). Left alone, such a network would eventually dominate the universe's energy budget, in conflict with observations. So for a domain-wall network to be cosmologically viable, it has to decay; the vacuum degeneracy needs to be lifted, whether explicitly or spontaneously, introducing a small pressure difference (or bias) $\Delta V$ between the two vacua or domains.

This bias exerts a volume pressure 
\begin{equation}
    p_V \simeq \Delta V,
\end{equation}
which pushes to shrink the higher-energy (false) vacuum regions, while the wall tension resists via
\begin{equation}
    p_T \simeq \frac{\sigma}{R},
\end{equation}
where $R$ denotes the typical curvature radius of the wall. During the scaling regime, the characteristic curvature radius is of the order of the Hubble radius \cite{hiramatsu2014estimation},
\begin{equation}
    R \sim H^{-1} \sim t,
\end{equation}
such that
\begin{equation}
    p_T \simeq \frac{\sigma}{t}.
\end{equation}

Initially, the tension pressure dominates, and the domain-wall network evolves toward the scaling solution. As the Universe expands, the tension pressure decreases while the volume pressure remains approximately constant. Domain-wall annihilation occurs once the two pressures become comparable, $p_T \simeq p_V$, which defines the annihilation time,
\begin{equation}
    t_{\rm ann} \simeq \frac{\sigma}{\Delta V}.
\end{equation}

It is convenient to parameterize the vacuum bias as
\begin{equation}
    \Delta V = \epsilon_b v_\phi^4,
\end{equation}
where $\epsilon_b \ll 1$ is a dimensionless bias parameter. The annihilation time can then be expressed as
\begin{equation}
    t_{\rm ann}
    \simeq
    \frac{f_\sigma}{\epsilon_b v_\phi},
\end{equation}
To be consistent with standard cosmology, the walls must annihilate before they can dominate the energy density. That collapse is violent; the energy locked in the network is released essentially all at once, and it generates a stochastic gravitational-wave background whose peak frequency and amplitude are set by $v_\phi$, $\sigma$, and $\Delta V$. The upper and lower bounds on the dimensionless bias parameter $\epsilon_b$ are found to be 
\begin{equation}
\label{eq:11e}
    10^{-25}\left(\frac{TeV}{v_\phi} \right) < \frac{\epsilon_b}{f_\sigma} < 10^{-15}\left(\frac{v_\phi}{TeV}\right)
\end{equation}
 which is discussed in  \cite{gelmini2021gravitational,chen2026gravitational}.

\section{\label{sec:citeref}MODEL FRAMEWORK}
In this work, we constructed a model based on $A_4 \otimes Z_4$ symmetry augmented by GCP symmetry 
to naturally generate the tri-bimaximal (TBM) mixing pattern at leading order while providing a predictive mechanism for the reactor angle $\theta_{13}$ and cosmological domain wall dynamics. To study the type-I seesaw mechanism for neutrino mass generation, we utilized the charged assignments for various fields listed in Table I. Under the $A_{4}$ symmetry, the three generations of the left-handed lepton doublet \(L\) are assigned to the triplet \(3\). Meanwhile, the right-handed charged leptons \(e_{R}\), \(\mu _{R}\), and \(\tau _{R}\) transform as the singlets \(1\), \(1^{\prime \prime }\), and \(1^{\prime }\), respectively. We extended the SM by introducing three heavy right-handed Majorana neutrinos $N_R$ considered as triplet 3 under $A_4$, alongside a specifically chosen flavon sector. The scalar sector comprises two $A_4$ triplets, $\phi$ and $\chi$ , and two $A_4$ singlets, $\eta$ and $\xi''$. Where $\eta$ is a trivial singlet 1 and $\xi''$ is a non-trivial singlet $1''$. 

\begin{table}[h!]
\centering
\renewcommand{\arraystretch}{1.4}
\small
\setlength{\tabcolsep}{5pt}
\caption{Field content of the model with their $A_4$ and $Z_4$ charges.}
\begin{tabular}{lccccccccccc}
\hline\hline
Symmetry & \multicolumn{9}{c}{Field content and charge} \\
\hline
& $L$ & $e_R$ & $\mu_R$ & $\tau_R$ & $N_{R}$ & $H$ & $\phi$ & $\chi$ & $\eta$ & $\xi''$ \\
\hline 
$A_4$ & 3 & 1 & $1''$ & $1'$ & 3 & 1 & 3 & 3 & 1 & $1''$ \\
$Z_4$ & $i$ & $i$ & $i$ & $i$ & $i$ & 1 & 1 & $-1$ & $-1$ & $-1$ \\
\hline\hline
\end{tabular}
\end{table}

In addition to the discrete flavor symmetries, the Lagrangian is mandated to be invariant under a GCP transformation. Under the action of the GCP operator, the fields transform as: $L \to U_r L^*$, $N_R \to U_r N_R^* $, $\phi \to U_r \phi^*$, $\chi \to U_r \chi^*$, $\eta \to \eta^*$, $\xi'' \to (\xi'')^*$ 
where $U_r = \mathbb{I}$ in the AF basis.

Since the fundamental Lagrangian is mandated to be exactly invariant under this GCP transformation, a rigid constraint is imposed on the parameter space: coupling constants associated with GCP-even operators must be strictly real, whereas those associated with GCP-odd operators are required to be purely imaginary to ensure the reality of the interaction terms.

Considering the charge assignments shown in Table I gives the charged lepton sector Lagrangian:
\begin{equation}
\begin{aligned}[b]
    \mathcal{L}_c={} & \frac{y_e}{\Lambda}[(\bar{L}\phi)_1 e_RH]+\frac{y_\mu}{\Lambda}[(\bar{L}\phi)_{1'} \mu_R H] \\
    & +\frac{y_\tau}{\Lambda}[(\bar{L}\phi)_{1''} \tau_R H]+h.c.
    \end{aligned}
\end{equation}
 From the \(A_{4}\) product rules, we obtain the following charged lepton mass matrix:
 \begin{equation}
     M_l=\begin{pmatrix}
     \frac{y_e v_\phi v_H}{\sqrt{2}\Lambda} & 0 & 0\\
     0 & \frac{y_\mu v_\phi v_H}{\sqrt{2}\Lambda} & 0\\
     0 & 0 & \frac{y_\tau v_\phi v_H}{\sqrt{2}\Lambda}
     \end{pmatrix}
 \end{equation}
  where, we have considered the VEV alignment $\langle \phi \rangle= v_\phi \begin{pmatrix}
      1 & 0 & 0
  \end{pmatrix}^T$.
 At leading order, the charged lepton mass matrix is diagonal, and GCP symmetry forces all $y_{e,\mu,\tau} \in \mathbb{R}$. 

 The invariant Dirac Lagrangian involving the lepton fields $L_i$ and heavy right-handed fields $N_{R_i}$ is of the form
 \begin{equation}
     \mathcal{L}_D=y_D[(\bar{L}N_R)_1 \tilde{H}]+h.c.
 \end{equation}
 The Dirac mass matrix is obtained to be 
 \begin{equation}
     M_D=\frac{y_D v_H}{\sqrt{2}}\begin{pmatrix}
         1 & 0 & 0\\
         0 & 0 & 1\\
         0 & 1 & 0
     \end{pmatrix}= \frac{y_D v_H}{\sqrt{2}}P_{23}
 \end{equation}
 Here, $\langle \tilde{H} \rangle=\frac{v_H}{\sqrt{2}}$ is the VEV of the Higgs field $H$ and $y_D \in \mathbb{R}$ as mandated by GCP.

The Heavy Majorana sector Lagrangian can be written as follows:
\begin{equation}
\begin{aligned}[b]
     \mathcal{L}_N= {} & \frac{y_\chi}{2}[(\bar{N_R^c}N_R)_{3_s}\chi]_{1}+\frac{y_\eta}{2}[(\bar{N_R^c}N_R)_{1} \eta]_{1} \\
     & + \frac{y_{\xi''}}{2}[(\bar{N_R^c}N_R)_{1'} \xi'']_{1}+ \frac{\tilde{y}_{\xi''}}{2}[(\bar{N_R^c}N_R)_{1''} (\xi'')^*]_{1} \\
     & +h.c.
\end{aligned}
\end{equation}

This gives the Majorana mass matrix of the form 
\begin{equation}
  M_N= \begin{pmatrix}
       2 b_1 + d_1 & -b_1 + c_1 & -b_1 + c_1\\
       -b_1 + c_1 & 2 b_1 + c_1 & -b_1 + d_1\\
       -b_1 + c_1 & -b_1 + d_1 & 2 b_1 + c_1
   \end{pmatrix} 
\end{equation}
Here, we have defined $b_1=\frac{y_\chi v_\chi}{2\sqrt{3}}$, $d_1=\frac{y_\eta v_\eta}{2}$ and $c_1=\frac{y_{\xi''} v_{\xi''}}{2}$ and considered the VEV alignments $\langle \chi \rangle=\frac{v_\chi}{\sqrt{3}}\begin{pmatrix}
    1 & 1 & 1
\end{pmatrix}^T$, $\langle \eta \rangle=v_\eta$ and $\langle \xi'' \rangle=v_{\xi''}$. Here, GCP also forced $y_{\chi,\eta,\xi''}$ to be real.

After integrating out the heavy right-handed neutrinos, the light neutrino mass is given by the type I seesaw mechanism as
\begin{equation}
    M_\nu=-M_D M_N^{-1}M_D^T
\end{equation}
The diagonalizing matrix of $M_N$ is found to be a tribimaximal mixing matrix, with the eigenvalues of $M_N$ of the form 
\begin{equation}
    m_{N_1}=3b_1 - c_1 + d_1, m_{N_2}= 2c_1 + d_1, m_{N_3}=3 b_1 + c_1 - d_1
\end{equation}
Since $M_N$ and $M_N^{-1}$ are diagonalized by the same unitary matrix $U_{\text{TBM}}$, with inverted eigenvalues, and given that $M_D = \frac{y_D v_H}{\sqrt{2}} P_{23}$ commutes with $P_{23}$ (where $P_{23}^2 = \mathbb{I}$), the resulting light neutrino mass matrix $M_\nu$ is likewise diagonalized by $U_{\text{TBM}}$. Consequently, with the charged-lepton mass matrix being diagonal in this basis, the PMNS mixing matrix takes the form,
\begin{equation}
    U_{PMNS}=U_l^\dagger U_\nu = U_\nu = U_{TBM}
\end{equation}
Thus,
\begin{equation}
    U_{TBM}^T M_\nu U_{TBM} = \frac{y_D^2 v_H^2}{2}diag(m_{N_1}^{-1},m_{N_2}^{-1},m_{N_3}^{-1})
\end{equation}
The experiments ruled out TBM mixing because it yielded $\theta_{13}=0$. Even though TBM mixing is disfavored, it can still serve as a viable leading-order approximation to the PMNS matrix. As discussed in \cite{chen2026gravitational}, incorporating all renormalizable terms in the flavon sector, including both the cubic couplings and the flavon cross-coupling terms, gives the VEVs of the flavon fields $\phi$ and $\chi$ non-trivial shifts (parameterized by $\epsilon_\phi$ and $\epsilon_\chi$). These cross-coupling-induced VEV perturbations lift the vacuum degeneracies and generate corrections to the lepton mass matrices, yielding mixing angles and a Dirac CP-violating phase that are fully consistent with the $3\sigma$ range of global neutrino oscillation data such as NuFIT \cite{esteban2024nufit,nufit61}. The calculation of this sort for the flavon sector has already been done in \cite{chen2026gravitational, pascoli2016flavon,pascoli2016role}.

In our model, the flavon sector analysis is almost similar to that of \cite{chen2026gravitational}; the only difference is the extra renormalizable terms containing $\xi''$, including cross-coupling terms between $\chi$ and $\xi''$, which connect the neutrino sector with the gravitational-wave sector through a common origin, which is the GCP-breaking ccross-coupling The form of all the renormalizable potentials of my model is shown in Appendix B, and also in Appendix D we showed that the contribution to the vacuum shift due to the introduction of the $\chi$ and $\xi''$ cross-coupling is zero. Assuming the perturbation to the VEV is small so that residual symmetries are preserved at leading order, the corrected VEV alignment for $\phi$ and $\chi$ can be written as 
\begin{equation}
\label{eq:22}
    \langle \phi \rangle=\begin{pmatrix}
        1\\
        \epsilon_\phi\\
        \epsilon_\phi
    \end{pmatrix}, \langle \chi \rangle=\begin{pmatrix}
        1-2\epsilon_\chi\\
        1+\epsilon_\chi\\
        1+\epsilon_\chi
        
    \end{pmatrix}
\end{equation}

\begin{widetext}
Where, 
\begin{equation}
    \epsilon_\phi=\frac{3v_\chi^2 \epsilon_2}{(2\sqrt{3}f_4-6(f_2 + f_3 +2(f_1+a(a^2+\sqrt{1+a^2})f_1+a(a+\sqrt{1+a^2})f_3))v_{\phi_-})v_{\phi_-}}
\end{equation}
and
\begin{equation}
    \epsilon_\chi=\frac{v_{\phi-}(3\sqrt{3}v_{\phi-}\epsilon_3 +2\epsilon_4)}{3(4\tilde{f_1}+\tilde{f_2}+3\tilde{f_3})v_\chi^2}
\end{equation}
Here, $\epsilon_2$, $\epsilon_3$, $\epsilon_4$, $f_1$, $f_2$, $f_3$, $f_4$ are the coupling constants of the flavon sector discussed in Appendix B, and GCP forces all these couplings to be real; hence the perturbation parameters $\epsilon_\phi$ and $\epsilon_\chi$ are also real, which is not the case in \cite{chen2026gravitational}. Since introducing cross-coupling between $\chi$ and $\xi''$ does not affect the shift, the forms of $\epsilon_\phi$ and $\epsilon_{\chi}$ remain unchanged from \cite{chen2026gravitational}.  

Considering this corrected VEV alignment shown in equation ~\eqref{eq:22} and the corrected VEV of $\langle \xi'' \rangle=v_{\xi''}e^{i\delta}$ (the VEV correction of $\xi''$ comes from the GCP-breaking quartic coupling, which is discussed in Appendix B and E), we get the charged lepton mass matrix of the form 
\begin{equation}
    M_l=\frac{v_H v_\phi}{\sqrt{2}\Lambda}\begin{pmatrix}
        y_e & \epsilon_\phi y_\mu & \epsilon_\phi y_\tau\\
        \epsilon_\phi y_e & y_\mu & \epsilon_\phi y_\tau\\
        \epsilon_\phi y_e & \epsilon_\phi y_\mu & y_\tau
    \end{pmatrix}
\end{equation}
and the Majorana mass matrix of the form
\begin{equation}
    M_N=\begin{pmatrix}
        2b_1+d_1-4b_1 \epsilon_\chi & -b_1+c_1 e^{i\delta}-b_1 \epsilon_\chi & -b_1+c_1 e^{-i\delta}-b_1 \epsilon_\chi\\
       -b_1+c_1 e^{i\delta}-b_1 \epsilon_\chi & 2b_1+c_1 e^{-i\delta}+2b_1\epsilon_\chi & -b_1+d_1+2b_1 \epsilon_\chi\\
       -b_1+c_1 e^{-i\delta}-b_1 \epsilon_\chi & -b_1 +d_1 +2b_1 \epsilon_\chi & 2b_1 +c_1 e^{i\delta}+2b_1\epsilon_\chi
    \end{pmatrix}
\end{equation}
This matrix is not diagonalizable by the TBM matrix; TBM gave 
\begin{equation}
    \tilde{M_N}=U_{TBM}^T.M_N.U_{TBM}=\begin{pmatrix}
        3b+d-c \cos{\delta} & -3\sqrt{2}b\epsilon_\chi & i\sqrt{3}c\sin{\delta}\\
        -3\sqrt{2}b\epsilon_\chi & d+2c\cos{\delta} & 0\\
        i\sqrt{3}c\sin{\delta} & 0 & 3b-d+c\cos{\delta}
    \end{pmatrix}
\end{equation}

Considering up to 1st order in perturbation, the TBM rotated light neutrino matrix is of the form 
\begin{equation}
    \tilde{M_\nu}=\begin{pmatrix}
       A & D\epsilon_\chi & iF\\
        D\epsilon_\chi & B & 0\\
       iF & 0 & C
    \end{pmatrix}
\end{equation}
Where,
\begin{equation}
\begin{aligned}
    A &= -\frac{3b_1-2c_1+d_1+c_1\cos{\delta}}{(3b_1-c_1+d_1)^2}=-\frac{m_{N_1}+c_1(\cos{\delta}-1)}{m_{N_1}^2}\\
    B &=-\frac{4c_1+d_1-2c_1\cos{\delta}}{(2c_1+d_1)^2}=-\frac{m_{N_2}+2c_1(1-\cos{\delta})}{m_{N_2}^2}\\
    C &=\frac{-3b_1-2c_1+d_1+c_1 \cos{\delta}}{(3b_1+c_1-d_1)^2}=-\frac{m_{N_3}+c_1(1-\cos{\delta})}{m_{N_3}^2}\\
    D &=\frac{-3\sqrt{2}b_1 \epsilon_\chi}{(3b_1-c_1+d_1)(2c_1+d_1)}=-\frac{3\sqrt{2}b_1}{m_{N_1}m_{N_2}}\\
    F &=-\frac{ \sqrt{3} c_1 \sin{\delta}}{-9b_1^2+(c_1-d_1)^2}=\frac{\sqrt{3}c_1 \sin{\delta}}{m_{N_1}m_{N_3}}\\
\end{aligned}
\end{equation}
\end{widetext}

After accounting for the corrected VEV alignment, the charge-lepton mass matrix becomes non-diagonal. Thus, diagonalizing the charged lepton mass matrix perturbatively using the relation $U_l^\dagger M_l M_l^\dagger U_l=diag(m_e^2,m_\mu^2,m_\tau^2)$, we get the diagonalizing matrix $U_l$ of the form 
\begin{equation}
    U_l^\dagger=\begin{pmatrix}
        1 & -\epsilon_\phi & -\epsilon_\phi\\
        \epsilon_\phi & 1 & -\epsilon_\phi\\
        \epsilon_\phi & \epsilon_\phi & 1
    \end{pmatrix}
\end{equation}
And we obtained the light neutrino mass matrix diagonalizing matrix $U_\nu$ in the form shown below via Takagi diagonalization. 
\begin{equation}
    U_\nu=\begin{pmatrix}
        1 & -\frac{3\sqrt{2}b_1\epsilon_\chi}{m_{N_2}-m_{N_1}} & \frac{i\sqrt{3}c_1 \sin{\delta}}{m_{N_1}+m_{N_3}}\\
        \frac{3\sqrt{2}b_1\epsilon_\chi}{m_{N_2}-m_{N_1}} &  1 & 0\\
        \frac{i\sqrt{3}c_1 \sin{\delta}}{m_{N_1}+m_{N_3}} & 0 & 1 
    \end{pmatrix}
\end{equation}
Thus the $U_{PMNS}$ matrix is of the form 
\begin{equation}
    U_{PMNS}=U_l^\dagger.U_{TBM}.U_\nu.P_\nu
\end{equation}
Here, the diagonal matrix incorporating the Majorana phase is denoted by $P_\nu$.

\section{NUMERICAL ANALYSIS}
\subsection{Neutrino Fit Data}
The three neutrino mixing angles are extracted from the PMNS matrix elements via the following relations:
\begin{equation}
    \begin{aligned}
    \label{eq:33}
        \sin^2{\theta_{13}} &=|U_{13}|^2 \approx \frac{2c_1^2\sin^2{\delta}}{36b_1^2}\\
        \sin^2{\theta_{12}} &=\frac{|U_{12}|^2}{1-|U_{13}|^2} \approx \frac{1}{3}(1-2\epsilon_\phi+\frac{2b_1\epsilon_\chi}{b_1-c_1})^2\\
        \sin^2{\theta_{23}} &=\frac{|U_{23}|^2}{1-|U_{13}|^2} \approx \frac{1}{2}(1+2\epsilon_\phi)
    \end{aligned}
\end{equation}

This gives TBM mixing in the limiting case. From equation ~\eqref{eq:33}, we can derive a sum rule between mixing angles as follows
\begin{equation}
    \sqrt{3}\sin{\theta_{12}}+2\sin^2{\theta_{23}}=2+\frac{2b_1\epsilon_\chi}{b_1-c_1}
\end{equation}
We also determine the Jarlskog invariant \(J_{CP}\) from the PMNS matrix elements via the relation below:
\begin{equation}
\begin{aligned}
    J_{CP} &=\mathrm{Im}[U_{11}U_{22}U_{12}^* U_{21}^*] \\
&=s_{12}c_{12}s_{23}c_{23}s_{13}c_{13}^2\sin{\delta_{CP}}
\end{aligned}
\end{equation}
here, $s_{ij}=\sin{\theta_{ij}}$ and $c_{ij}=\cos{\theta_{ij}}$.\\

For our numerical calculations, we incorporate global fit data on neutrino oscillations at the 3$\sigma$ level, as outlined in Table II. We have defined the dimensionless parameters $c=\frac{c_1}{b_1}$ and $d=\frac{d_1}{b_1}$ and considered the following parameter ranges to ensure compatibility with neutrino oscillation data.
\begin{equation}
\begin{aligned}
    c & \in [0.63, 0.94], \quad d \in [1.967, 2.594], \\
   &\epsilon_\phi \in [-0.061, 0.060], \quad \epsilon_\chi \in [-0.024, 0.01], \\
   &\delta_{GCP} \in [42.63^0, 137.36^0], \quad m_0 \in [0.0537, 0.1134]
\end{aligned}
\end{equation}
This scan is restricted to normal ordering (NO), the only mass ordering compatible with the model, as no (c,d) satisfying the above ranges yields a physical inverted-ordering spectrum.
Here, $m_0 =\frac{y_D^2v_H^2}{2b_1}$.

Within these designated parameter ranges, the input parameters are scanned at random. To determine the permitted parameter regions, we employ the \(3\sigma\) constraints on mixing angles and mass-squared differences listed in Table II. The correlation between parameters $c$ and $d$ is shown in the left panel of Fig. 1, and the right panel of Fig. 1 shows the correlation between the perturbation parameters $\epsilon_\phi$ and $\epsilon_\chi$. 

In Fig. 2, we have shown the correlation between the neutrino mixing angles $\sin^2{\theta_{13}},\sin^2{\theta_{12}},\sin^2{\theta_{23}}$ and the sum of neutrino masses $\sum{m_\nu}$. Our results yield a total neutrino mass that remains well within the upper limit $\sum{m_\nu}=120$ meV set by Planck data \cite{aghanim2018planck}. 

In the left panel of Fig. 3, we have shown the correlation of the Jarlskog invariant ($J_{CP}$) with the sum of the neutrino masses $\sum{m_\nu}$. The $J_{CP}$ is found to be in the range [-0.0380799, -0.0323949], and in the right panel of Fig. 3, we have shown the correlation between $J_{CP}$ and $\delta_{CP}$ and we found that for allowed range of $J_{CP}$ the CP phase $\delta_{CP}$ is $270^0$.

In Fig. 4, we have shown the correlation between $\sin^2{\theta_{13}}$ and $J_{CP}$. 

\begin{widetext}
\begin{figure*}[htbp]
 \centering
\begin{minipage}{0.48\linewidth}
\centering
\includegraphics[width=\linewidth]{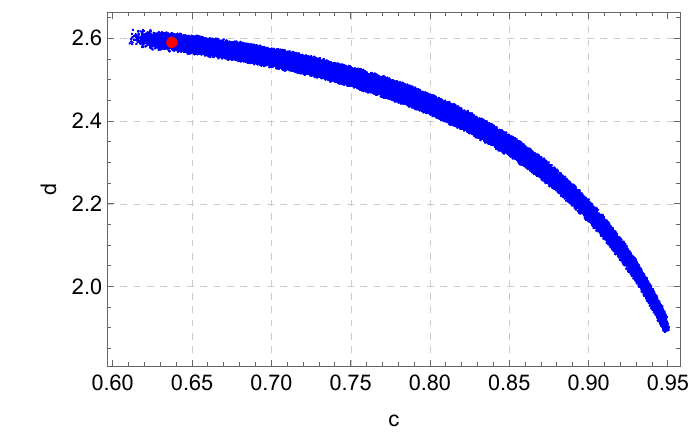}
     \label{fig:cd}
     \end{minipage}
     \hfill 
    \begin{minipage}{0.48\linewidth}
       \centering
       \includegraphics[width=\linewidth]{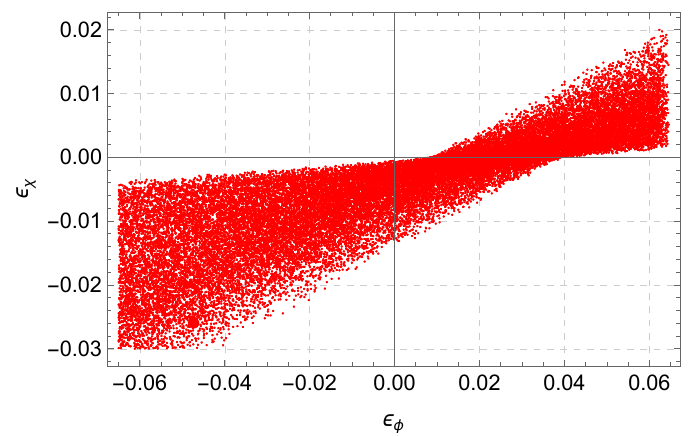}
       \label{fig:ep}
    \end{minipage}
\caption{Left: Correlation between model parameters c and d. Right: Correlation between perturbation term $\epsilon_{\phi}$ and $\epsilon_{\chi}$.}

    \label{fig:top_row_correlation}

\end{figure*}

\begin{figure*}[htbp]
    \centering
   
    \begin{minipage}{0.48\linewidth}
        \centering
        \includegraphics[width=\linewidth]{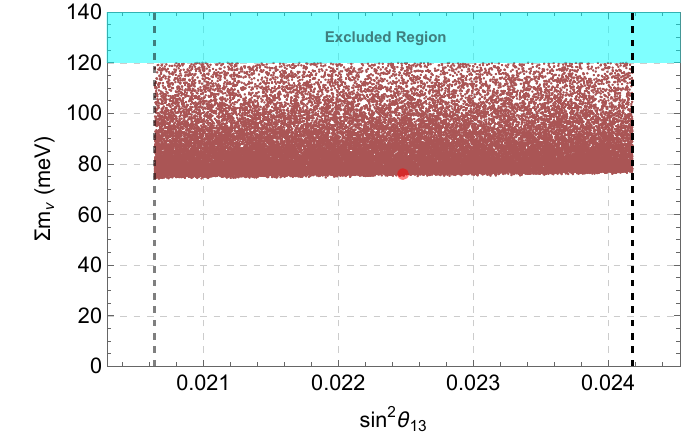}
        \label{fig:sin13}
    \end{minipage}
    \hfill 
    \begin{minipage}{0.48\linewidth}
        \centering
        \includegraphics[width=\linewidth]{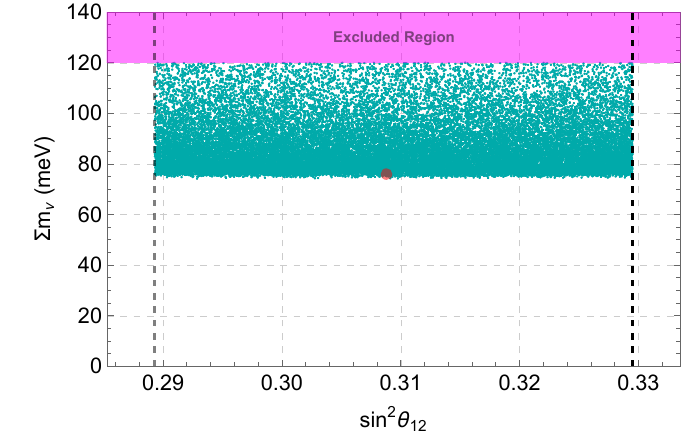}
        \label{fig:sin12}
    \end{minipage}
    
    \vspace{2ex} 
    
    \begin{minipage}{0.48\linewidth}
        \centering
        \includegraphics[width=\linewidth]{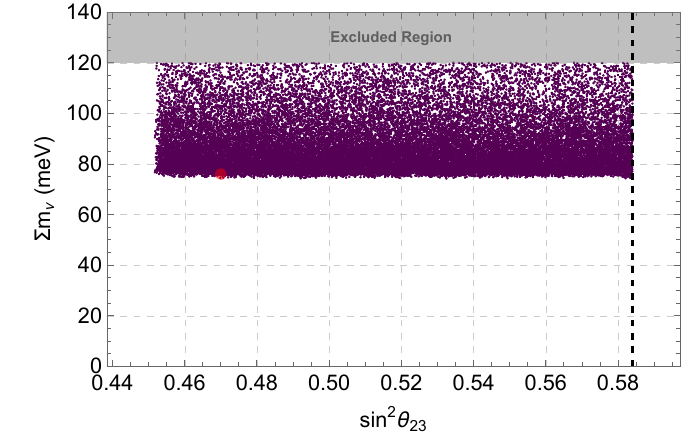}
        \label{fig:sin23}
    \end{minipage}

    \caption{Correlations between the mixing angles ($\sin^2\theta_{13}$, $\sin^2\theta_{12}$, $\sin^2\theta_{23}$) and the total neutrino mass $\sum m_{\nu}$.}
    \label{fig:mixing_angles_correlation}
\end{figure*}

\begin{figure*}[htbp]
    \centering
    \begin{minipage}{0.48\linewidth}
        \centering
        \includegraphics[width=\linewidth]{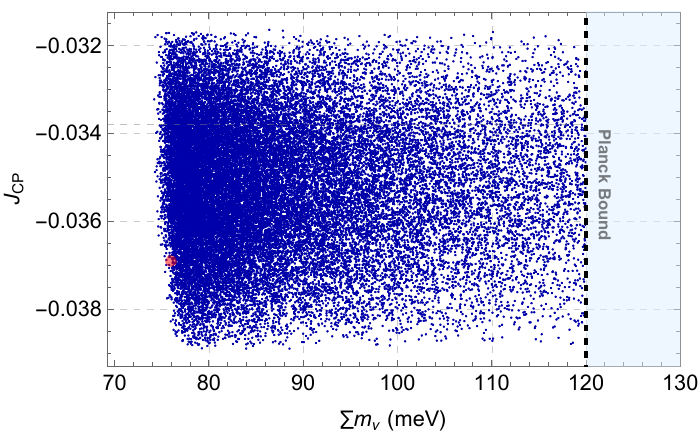}
        \label{fig:Jcp}
    \end{minipage}
    \hfill 
    \begin{minipage}{0.48\linewidth}
        \centering
        \includegraphics[width=\linewidth]{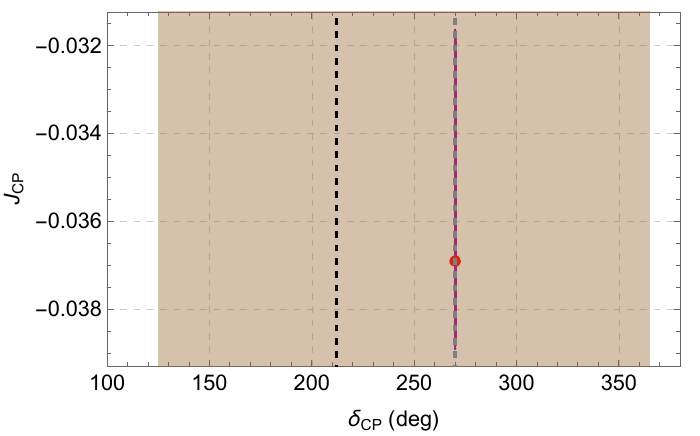}
        \label{fig:Jcp delta}
    \end{minipage}
    
    \caption{Left: Corellation between Jarlskog invariant $J_{CP}$ and sum of neutrino mass $\sum m_{\nu}$. Right: Correlation between $J_{CP}$ and $\delta_{CP}$(black dashed line represents best fit value of $\delta_{CP}$ ) and coloured region represents NuFIT $3\sigma$ range for $\delta_{CP}$.}
    \label{fig:top_row_correlation}
\end{figure*}

\begin{figure}[htbp]
    \centering
    \includegraphics[width=0.48\linewidth]{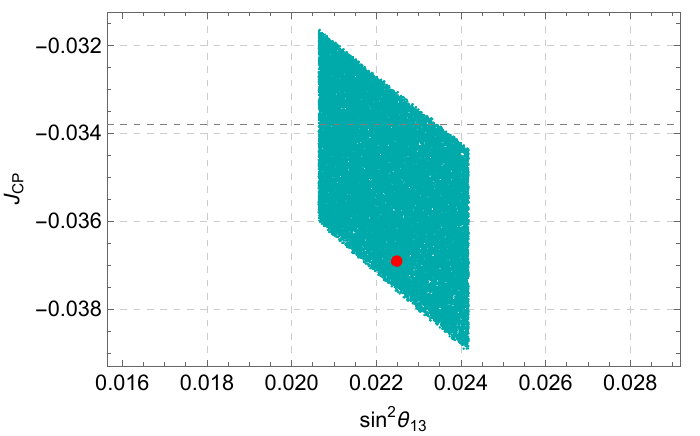}
    \caption{Correlation of $J_{CP}$ with $\sin^2{\theta_{13}}$.}
    \label{fig:placeholder}
\end{figure}
\end{widetext}

\begin{table}[htbp]
\caption{The NuFIT 6.1 (2025) results\cite{esteban2024nufit,nufit61}.}
\centering
\begin{tabular*}{\linewidth}{l @{\extracolsep{\fill}} cc}
\hline\hline
& \multicolumn{2}{c}{Normal ordering} \\
\hline
& $\text{bfp} \pm 1\sigma$ & $3\sigma \text{ range}$ \\[6pt]
\hline
$\sin^2 \theta_{12}$ & $0.3088^{+0.0067}_{-0.0066}$ & $0.2893-0.3295$ \\[6pt]
$\sin^2 \theta_{13}$ & $0.02248^{+0.00055}_{-0.00059}$ & $0.02064-0.02418$ \\[6pt]
$\sin^2 \theta_{23}$ & $0.470^{+0.017}_{-0.014}$ & $0.435-0.584$ \\[6pt]
$\dfrac{\Delta m_{21}^2}{10^{-5}\,\text{eV}^2}$ & $7.537^{+0.094}_{-0.10}$ & $7.236-7.823$ \\[6pt]
$\dfrac{\Delta m_{31}^2}{10^{-3}\,\text{eV}^2}$ & $2.511^{+0.021}_{-0.020}$ & $2.450-2.576$ \\[6pt]
$\delta_{CP}/^\circ$ & $212^{+26}_{-36}$ & $125-365$ \\[6pt]
\hline\hline
\end{tabular*}
\end{table}

\subsection{Neutrinoless double beta decay ($\nu0\beta\beta$)}
Neutrinoless double beta decay is a hypothetical, rare, lepton-number-violating nuclear transition of the type $(A, Z) \rightarrow (A, Z+2) + 2e^-$. The observation of neutrinoless double-beta decay, a process where a nucleus decays without neutrino emission, would prove that neutrinos are Majorana fermions. Furthermore, it would confirm that the total lepton number is violated by two units (\(\Delta L = 2\)).

The effective mass $ \left| m_{ee} \right|$, which encapsulates the fundamental particle physics parameters (The decay width of the $\nu 0 \beta \beta$ decay is proportional to $\left| m_{ee}\right|$) constrained by neutrino oscillation experiments, is defined as:
\begin{equation}
    m_{ee} = \left| \sum_{i=1}^{3} U_{ei}^2 m_i \right|
\end{equation}
Expanding this expression in terms of the standard parametrization of the PMNS mixing matrix elements $U_{ei}$, the neutrino mass eigenvalues $m_i$ ($i = 1, 2, 3$), and the Majorana CP-violating phases yields:
\begin{equation}
\begin{split}
    \left|m_{ee}\right| = \Big| & m_1 \cos^2{\theta_{12}}\cos^2{\theta_{13}} \\
    & + m_2 \sin^2{\theta_{12}}\cos^2{\theta_{13}} e^{i\alpha_1} \\
    & + m_3 \sin^2{\theta_{13}} e^{i(\alpha_2 - 2\delta_{CP})} \Big|
\end{split}
\end{equation}
In Fig. 5, we have shown the correlation between the effective neutrino mass $|m_{ee}|$ and the sum of the neutrino masses $\sum{m_\nu}$. We found that model predictions lie well within the sensitivity reach of upcoming experiments such as nEXO \cite{adhikari2022nexo}, which is in the range $(5-20)$ meV at $90\%$ CL, and the KamLAND-Zen experiment \cite{abe2025search}, which lies within $(28-122)$ meV.
\begin{figure}[htbp]
    \centering
    \includegraphics[width=1\linewidth]{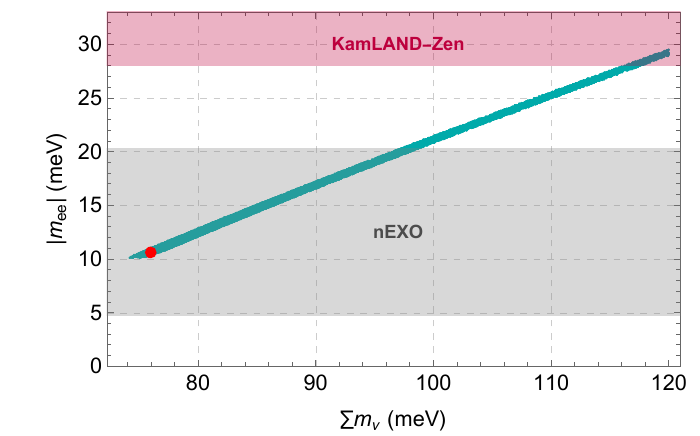}
    \caption{The effective Majorana neutrino mass \(\vert{}m_{ee}\vert{}\) plotted against the total neutrino mass \(\sum m_{\nu}\). The shaded bands indicate the sensitivity reach of current experiments.}
    \label{fig:placeholder}
\end{figure}

\section{Gravitational Waves from Domain Wall Annihilation}
As we discussed in Section II, due to bias, the domain walls get annihilated, and this produces gravitational waves. The gravitational wave spectrum emitted due to the domain wall collapse at time $t$ with frequency $f$ can be calculated using 
\begin{equation}
    \Omega h^2(f,t)=\frac{h^2}{\rho_c(t)}\frac{d\rho_{GW}(t)}{d\ln f}
\end{equation}
And its peak amplitude and peak frequency, red-shifted to the present day, can be calculated using the following equations
\begin{equation}
\label{eq:40}
    \Omega h^2|_{peak} \approx 10^{-67}\frac{f_\sigma^4}{\epsilon_b^2}\left(\frac{v}{TeV}\right)^4
\end{equation}
and, 
\begin{equation}
\label{eq:41}
    f_{peak} \approx 3 \times 10^3 Hz \left(\frac{\epsilon_b v}{f_b \text{TeV}}\right)^{\frac{1}{2}}
\end{equation}

Here, $\epsilon_b$ is the bias parameter shown in equation (9). Our model incorporates two independent sources of vacuum bias that lift the degeneracy of the $Z_3$ and $ Z_2$-preserving vacua. The first source stems from the cubic coupling of $\phi$, which generates a bias among the $ Z_3$-preserving vacua (discussed in Appendix C), while the second arises from the GCP-violating potential term involving $\chi$ and $\xi''$, which splits the $ Z_2$-preserving vacua (derived in Appendix E). The bias parameters are 
\begin{equation}
\begin{aligned}
\label{eq:42}
    \epsilon_b^{z_3} &= \frac{4a(3g_1+2g_2)(1+a^2)^{3/2}}{(\sqrt{1+a^2}-a)^4} \\
    \epsilon_b^{Z_2} &=
\frac{3\left|\kappa\right|v_{\xi''}^2}{v_\chi^2}
    \end{aligned}       
\end{equation}

We assumed a broken power law that complies in order to plot the gravitational wave spectrum as discussed in \cite{hiramatsu2014estimation}
\begin{equation}
    \begin{aligned}
        \Omega h^2 &\propto f^3  \quad \text{for}\quad f< f_{peak}\\
        &\propto f^{-1} \quad\text{for} \quad f>f_{peak}
    \end{aligned}
\end{equation}
Assuming a smooth broken power law, as in \cite{notari2025spectrum}, we have plotted the gravitational-wave spectrum in Fig. 6.
\begin{widetext}
   \begin{figure*}[htbp]
    \centering
   
    \begin{subfigure}{0.48\linewidth}
        \centering
        \includegraphics[width=\linewidth]{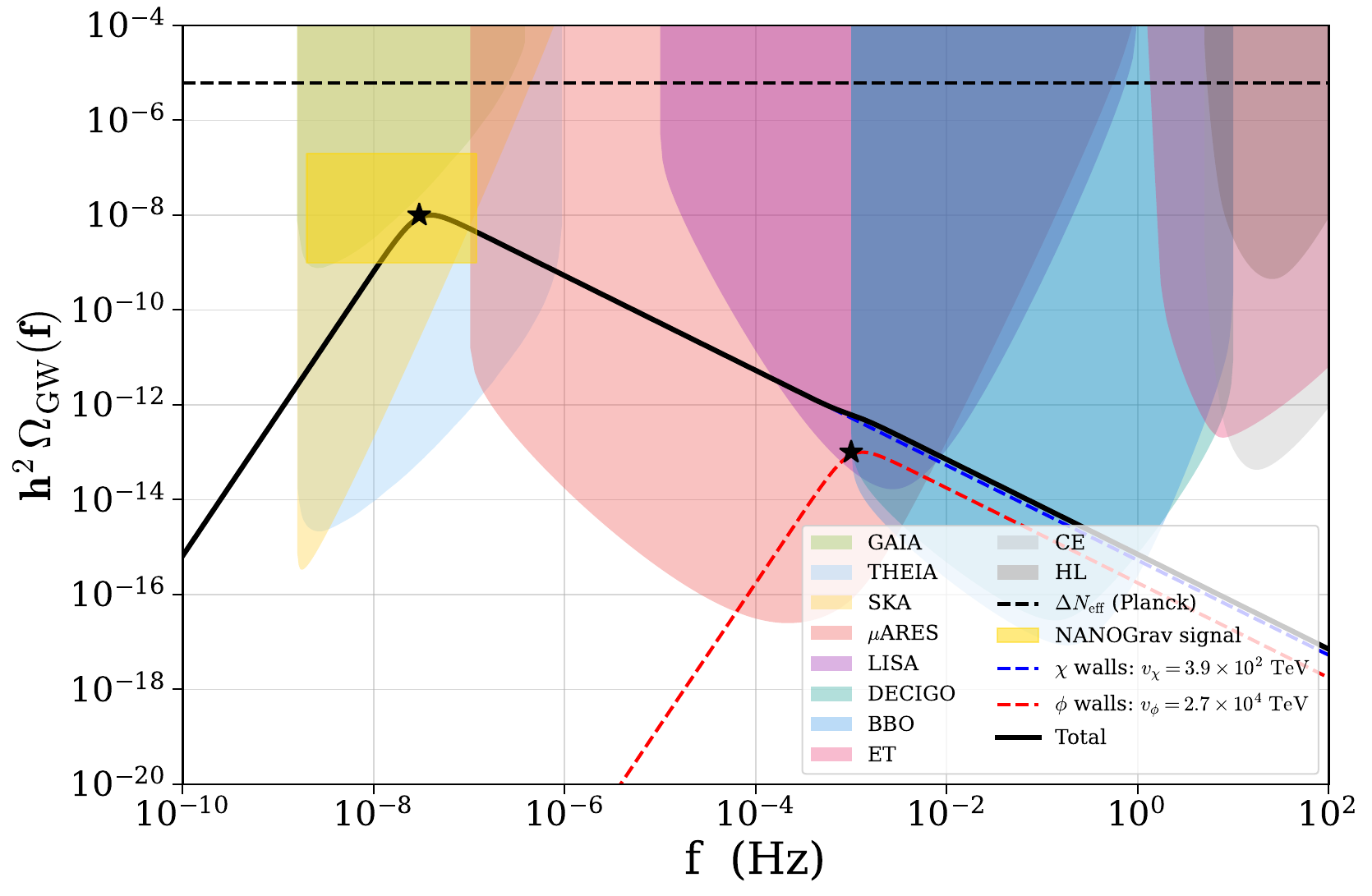}
        \caption{} 
        \label{fig:GWtotal}
    \end{subfigure}
    
    \vspace{2ex} 
    
    \begin{subfigure}{0.48\linewidth}
        \centering
        \includegraphics[width=\linewidth]{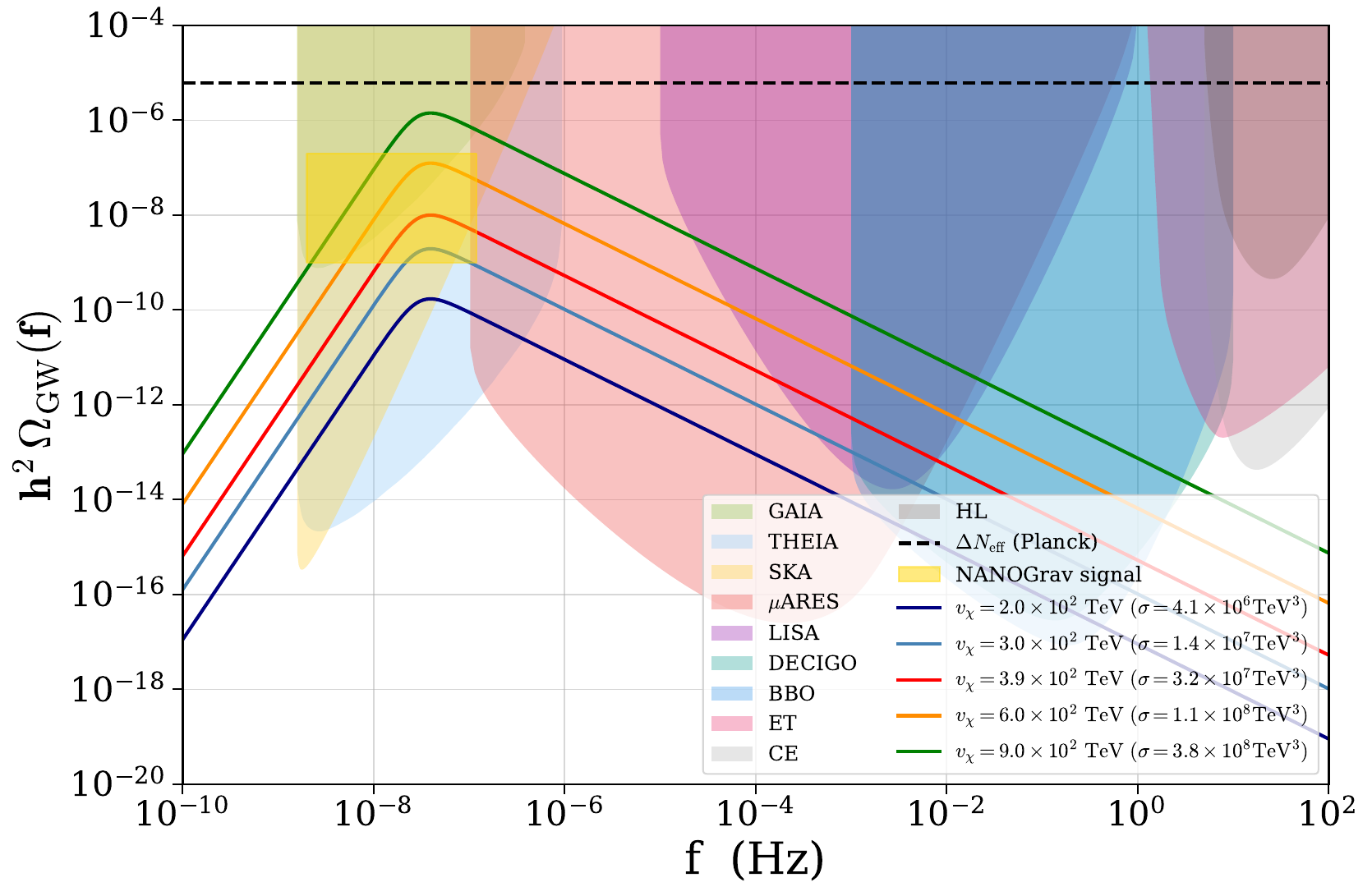}
        \caption{} 
        \label{fig:GWchi}
    \end{subfigure}
     \hfill
    \begin{subfigure}{0.48\linewidth}
        \centering
        \includegraphics[width=\linewidth]{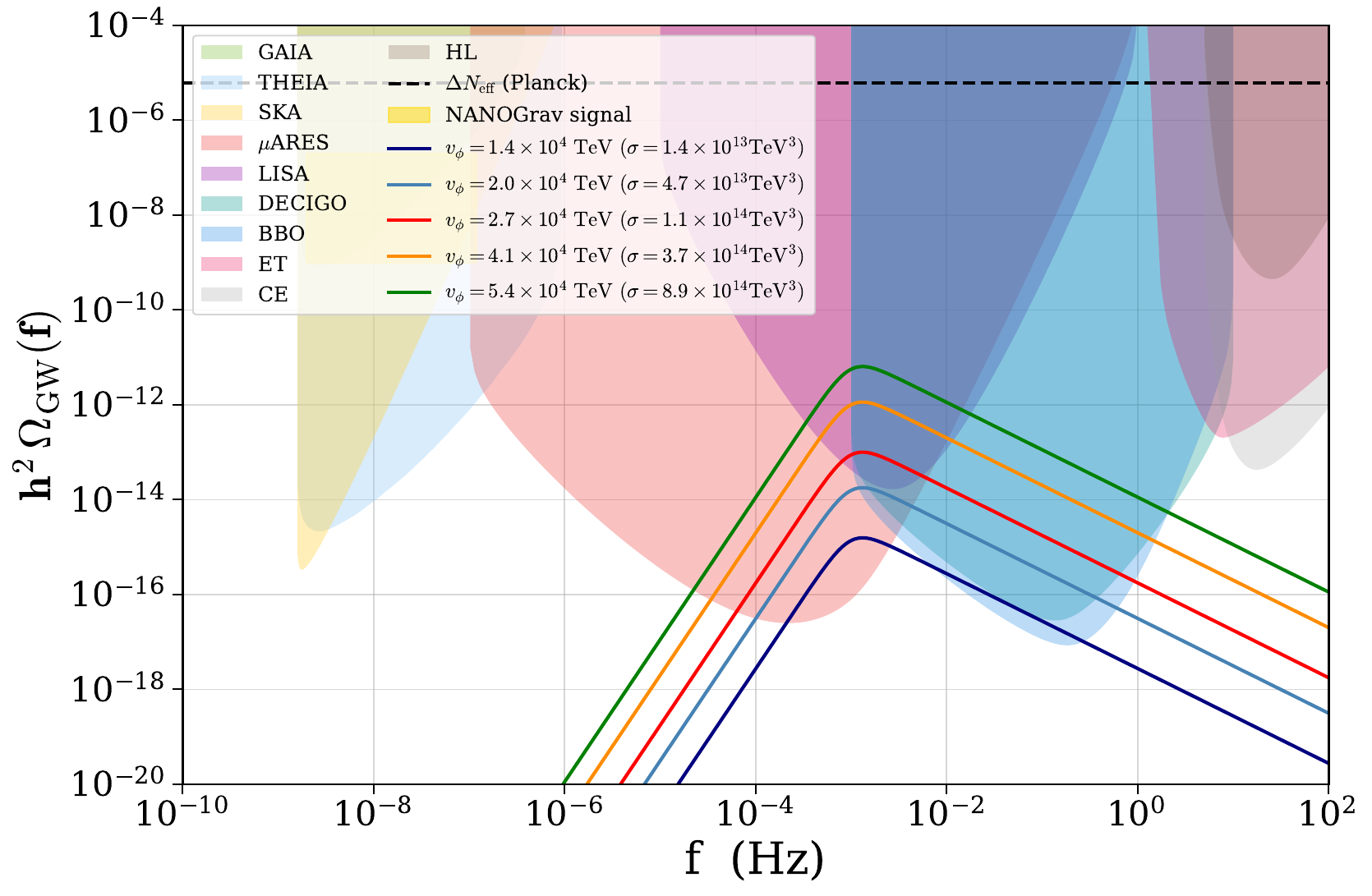}
        \caption{} 
        \label{fig:GWphi}
    \end{subfigure}

    \caption{(a) Resultant GW spectrum from domain-wall collapse. (b) GW spectrum for different $\chi$ VEV ($Z_2$ wall tension) values. (c) GW spectrum for different $\phi$ VEV ($Z_3$ wall tension) values.}
    \label{fig:mixing_angles_correlation}
\end{figure*}
\end{widetext}

Also, using Eqs. ~\eqref{eq:33} and ~\eqref{eq:42}, we get a very interesting relation between $\theta_{13}$ and the $Z_2$ bias term $\epsilon_b^{\xi''}$ (the detailed derivation is shown in Appendix E), shown below
\begin{equation}
 \epsilon_b^{\xi''}=\mathcal{K}\sin^2{\theta_{13}}
\end{equation}
Combining this relation with the peak amplitude and frequency formulas of Eq. ~\eqref{eq:40} and ~\eqref{eq:41}, we obtain a gravitational-wave spectrum determined directly by the measured value of the reactor angle, as shown in Fig. 7.
\begin{widetext}
\begin{figure*}[htbp]
    \centering
    \begin{minipage}{0.48\linewidth}
        \centering
        \includegraphics[width=\linewidth]{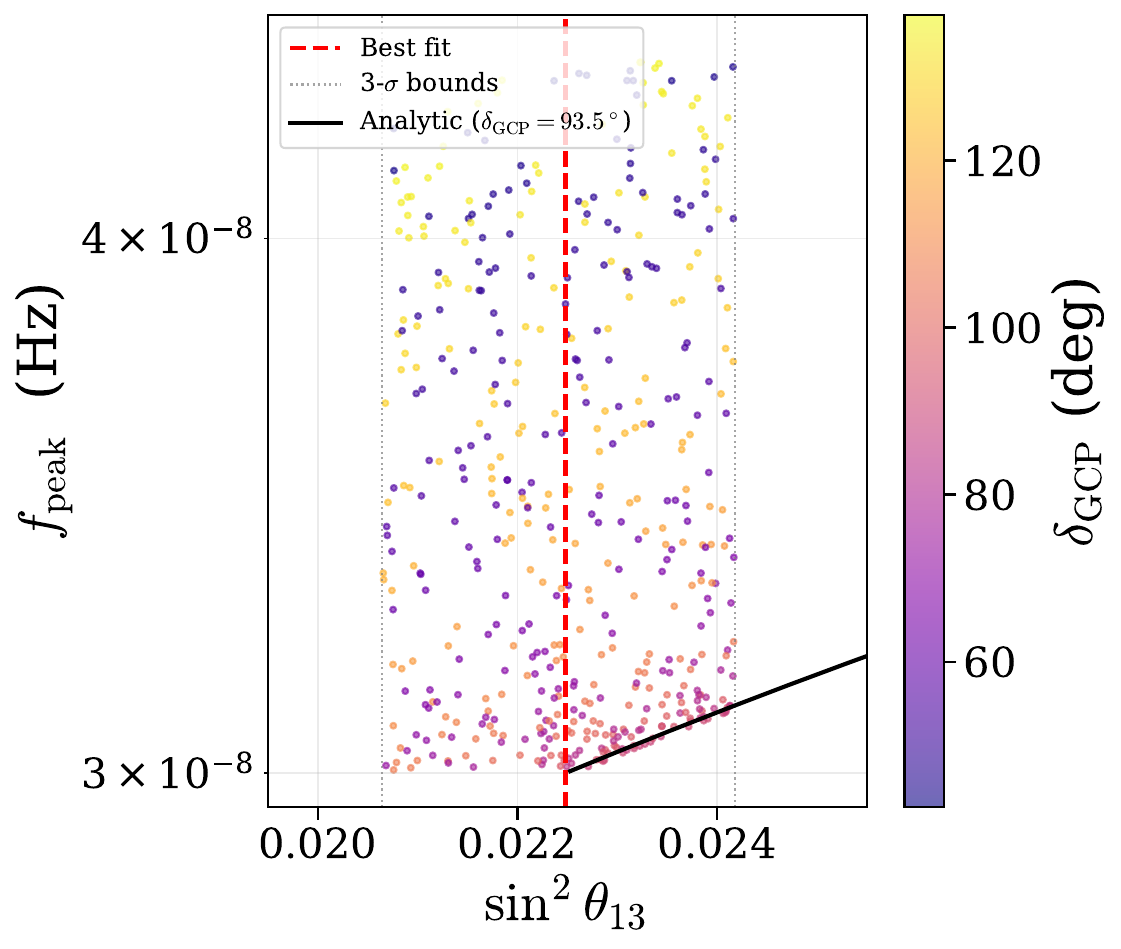}
        \label{fig:th13a}
    \end{minipage}
    \hfill 
    \begin{minipage}{0.48\linewidth}
        \centering
        \includegraphics[width=\linewidth]{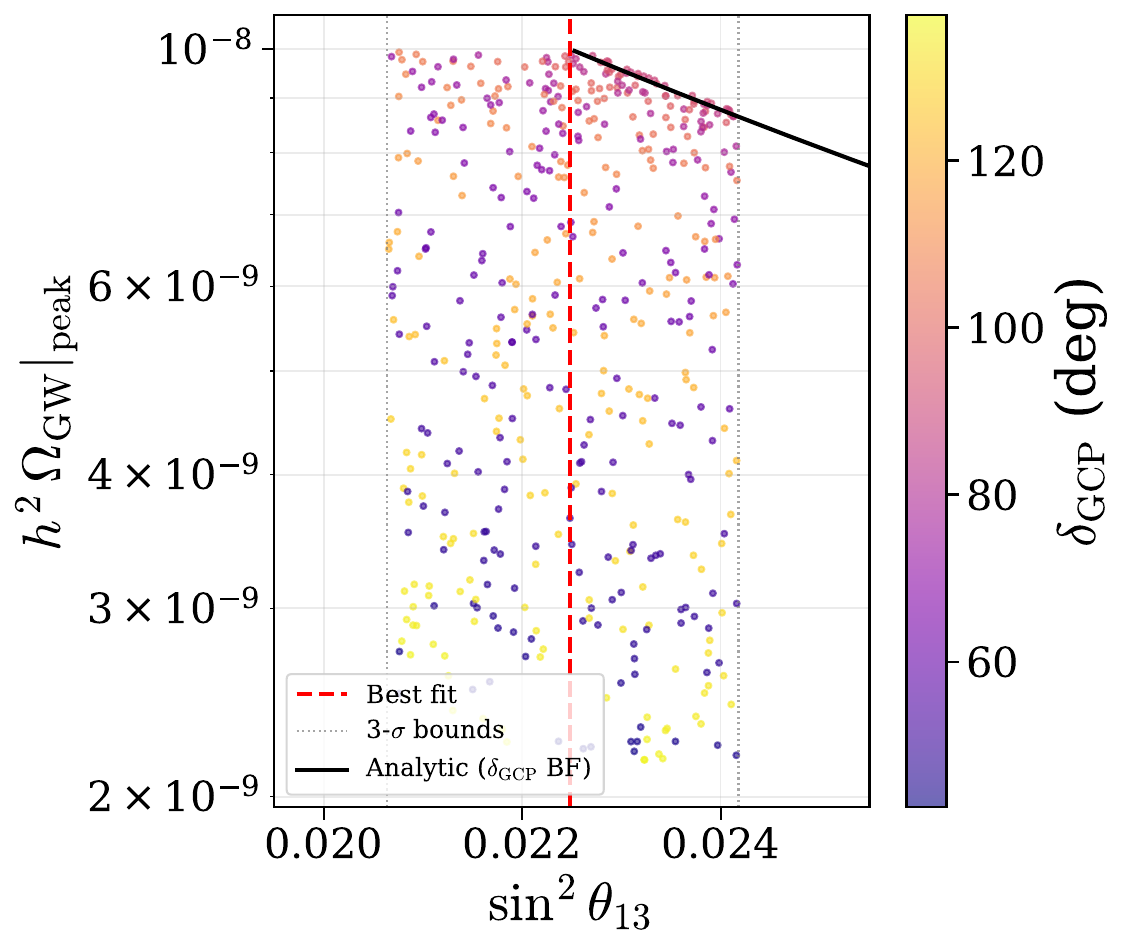}
        \label{fig:th13b}
    \end{minipage}
    
    \caption{Left: Correlation between peak frequency and $\theta_{13}$. Right:  Correlation between peak amplitude and $\theta_{13}$.}
    \label{fig:top_row_correlation}
\end{figure*}
\end{widetext}

\section{Conclusion}

In this work, we have constructed an $A_4 \times Z_4$ flavor model with generalized CP symmetry in which a single new operator, involving the $A_4$ singlet flavon $\xi''$, simultaneously explains the observed reactor mixing angle and sources a detectable stochastic gravitational-wave background. This dual role follows directly from the structure of $V_{\xi''\chi} = \kappa(\xi'')^2(\chi\chi)_{1''} + \text{h.c.}$, the GCP-violating phase $\phi_\kappa = \arg(\kappa)$ fixes via $\delta = (\pi - \phi_\kappa)/2$ and thereby $\sin\theta_{13}$, while this same coupling sets the bias energy shown in the Eq. ~\eqref{eq:42} that drives the annihilation of $\mathbb{Z}_2$ domain walls.

The $\kappa$ is the unique coupling whose phase cannot be removed by any field redefinition consistent with the rest of the Lagrangian, since the residual rephasing freedom of $\xi''$ is entirely fixed by its appearance in the heavy Majorana sector. As a result, $\mathrm{Im}(\kappa) \neq 0$ is the sole, explicit source of CP violation in the model, a single physical parameter responsible for two a priori unrelated phenomena. Its phase, $\phi_\kappa = \arg(\kappa)$, is protected at two GCP-conserving fixed points, $\delta = 0$ and $\delta = \pi/2$; our global fit selects the latter, at which the reactor angle is set predominantly by the mass-sector parameter $c$ while the Dirac phase is pinned to its maximal, symmetry-protected value $\delta_{\text{CP}} = 270^\circ$, with $\phi_\kappa$ controlling the small residual deviation from this point. Its magnitude depends on the choice of flavon scale. The residual $\mathbb{Z}_2$ degeneracy of the $\chi$ vacua is restored exactly as $\kappa \to 0$.

Our global fit to NuFit 6.1 \cite{esteban2024nufit,nufit61} data, performed in the Altarelli--Feruglio basis, yields $c = 0.6373$, $d = 2.5906$, $\delta_{\text{GCP}} = 93.53^\circ$, and $m_0 = 53.7$ meV, reproducing all measured oscillation parameters at best fit while predicting $\delta_{\text{CP}} = 270^\circ$, $J_{CP}=[-0.032,-0.039]$ and $|m_{ee}| = 14.6^{+12.7}_{-4.0}$ meV. These predictions depend only on the dimensionless ratios $c, d, \varepsilon_\phi, \varepsilon_\chi, \delta_{GCP}$ and are therefore completely insensitive to the absolute flavon-symmetry-breaking scale, leaving the neutrino sector unaffected by any subsequent choice made in the gravitational-wave analysis.

Setting the cross-coupling parameter $\varepsilon_4 = 0$, a choice protected by the enhanced symmetry it restores, and one that leaves $\varepsilon_\phi, \varepsilon_\chi$ (sourced instead by $\varepsilon_2, \varepsilon_3$) fully intact and isolates $\xi''$ as the only source of $\mathbb{Z}_2$ bias. Demanding that the domain-wall tension be computed self-consistently as $\sigma = f_\sigma v^3$ from the same perturbative quartic couplings used throughout, rather than fit as an independent free parameter, fixes $v_\chi \simeq 394$ TeV and $v_\phi \simeq 2.7 \times 10^4$ TeV, requiring a Dirac Yukawa $y_D \sim 2 \times 10^{-5}$, comparable to the electron Yukawa and thus fully natural for a type-I seesaw at this scale. At this benchmark, the $\chi$-wall spectrum peaks at $f \sim 3 \times 10^{-8}$ Hz with $h^2\Omega_\text{GW} \sim 10^{-8}$, within the pulsar-timing-array band \cite{antoniadis2023second}, while the $\phi$-wall spectrum peaks at $f \sim 10^{-3}$ Hz within the LISA band \cite{colpi2024lisa}; both satisfy the cosmological viability bound on the bias-to-tension ratio of Eq. ~\eqref{eq:11e}. 
The distinctive prediction of our model is the direct correlation between the reactor mixing angle and the gravitational-wave spectrum: $f_{\text{peak}} \propto \sin\theta_{13}$ and $\Omega_{\text{peak}} \propto \sin^{-4}\theta_{13}$, both traceable to the single complex coupling. Future improvements in reactor and long-baseline measurements of $\theta_{13}$ will sharpen this prediction. At the same time, forthcoming pulsar-timing-array and LISA data offer an independent, complementary test, establishing a rare observational bridge between low-energy neutrino physics and gravitational-wave cosmology.

\section*{Appendix A}
$A_4$ has four irreducible representations $1$, $1'$, $1''$ and $3$. The $A_4$ tensor product rules we have used are shown below 
\begin{equation}
\begin{aligned}
    1 \otimes 1 &=1\\
    1' \otimes 1' &=1''\\
    1' \otimes 1'' &=1\\ 
    1'' \otimes 1'' &=1'\\
    1^{(')('')} \otimes 3 &=3\\
    3 \otimes 3 &= 1\oplus 1' \oplus 1'' \oplus 3_S \oplus 3_A  
\end{aligned}
\end{equation}
If $a=(a_1,a_2,a_3)$ and $b=(b_1,b_2,b_3)$ are two triplets of $A_4$, the tensor product of them in the Altarelli-Feruglio (AF) \cite{altarelli2006tri} basis is of the following form
\begin{equation}
    \begin{aligned}
        (ab)_1 &=a_1b_1 + a_2b_2+a_3b_3\\
        (ab)_{1'} &= a_3b_3+a_1b_2+a_2b_1\\
        (ab)_{1''} &= a_2b_2+a_3b_1+a_1b_3\\
    \end{aligned}
    \end{equation}
\begin{equation}
        (ab)_{3_S} = 
        \frac{1}{2} 
        \begin{pmatrix}
            2a_1b_1-a_2b_3-a_3b_2\\
            2a_3b_3-a_1b_2-a_2b_1\\
            2a_2b_2-a_3b_1-a_1b_3
        \end{pmatrix}
\end{equation}
\begin{equation}
        (ab)_{3_A} = 
        \frac{1}{2}
        \begin{pmatrix}
            a_2b_3-a_3b_2\\
            a_1b_2-a_2b_1\\
            a_3b_1-a_1b_3
        \end{pmatrix}
\end{equation}
The tensor product of them in the Ma-Rajasekaran (MR) \cite{ma2001softly} basis is as follows
\begin{equation}
    \begin{aligned}
        (ab)_1 &=a_1b_1+a_2b_2+a_3b_3\\
        (ab)_{1'} &=a_1b_1+\omega a_2b_2+ \omega^2 a_3b_3\\
        (ab)_{1''} &=a_1b_1+\omega^2a_2b_2+\omega a_3b_3
    \end{aligned}
\end{equation}
\begin{equation}
     (ab)_{3_S} = \frac{\sqrt{3}}{2}\begin{pmatrix}
            a_2b_3+a_3b_2\\
            a_3b_1+a_1b_3\\
            a_1b_2+a_2b_1
        \end{pmatrix}
\end{equation}
    \begin{equation}
        (ab)_{3_A} = \frac{i}{2}\begin{pmatrix}
            a_2b_3-a_3b_2\\
            a_3b_1-a_1b_3\\
            a_1b_2-a_2b_1
        \end{pmatrix}
    \end{equation} 
Here, $\omega=e^{2\pi i/3}$

\section*{Appendix B}
To establish the vacuum structure of the model, we consider the most general renormalizable scalar potential invariant under the $A_4 \otimes Z_4 \otimes GCP$ flavor symmetry. The flavon sector comprises the triplet fields $\phi$ and $\chi$, as well as $\eta$ and $\xi''$. The potential can be decomposed into individual contributions and their cross-interactions:
\begin{widetext}
\begin{equation}
    V=V(\phi)+V(\chi)+V(\eta)+V(\xi'')+V_{cross}+ V(\chi \xi'')
\end{equation}
For the flavon $\phi$, which transforms as a triplet under $A_4$, carries a $+1$ charge under $\mathbb{Z}_4$ and GCP even, the invariant potential is given by 
\begin{equation}
\label{eq:53}
    V(\phi)=\frac{1}{2}\mu_\phi^2(\phi \phi)_1 + \frac{1}{4}[f_1(\phi \phi)_1^2+f_2(\phi \phi)_{1'}(\phi \phi)_{1''}+f_3((\phi \phi)_{3s}(\phi \phi)_{3s})_1]+\frac{f_4}{3}((\phi \phi)_{3s}\phi)_1
\end{equation}
Here, $\mu_\phi^2$ is mass like term and $f_1$,$f_2$,$f_3$ and $f_4$ are free couplings.

Similarly, the potential for $\chi$ (carrying a $-1$ charge under $\mathbb{Z}_4$ and GCP even) takes an identical structural form, characterized by parameters $\tilde{\mu}_\chi^2$ and $\tilde{f}_i$, but strictly forbids a cubic term due to its $\mathbb{Z}_4$ assignment
\begin{equation}
    V(\chi)=\frac{1}{2}\mu_\chi^2(\chi \chi)_1 + \frac{1}{4}[\tilde{f_1}(\chi \chi)_1^2 + \tilde{f_2}(\chi \chi)_{1'}(\chi \chi)_{1''}+\tilde{f_3}((\chi \chi)_{3s} (\chi \chi)_{3s})_1]
\end{equation}
Here also, $\mu_\chi^2$ is mass like term and $\tilde{f_1}$,$\tilde{f_2}$ and $\tilde{f_3}$ are free couplings.

The potential for $\eta$ ($A_4$ singlet 1 and carring $Z_4$ charge -1) and $\xi''$ ($A_4$ singlet $1''$ and carring $Z_4$ charge -1) are of the form
\begin{equation}
    V(\eta)=-\mu_\eta^2\left|\eta\right| + \lambda_\eta \left|\eta\right|^4
\end{equation}

\begin{equation}
    V(\xi'')=-\mu_{\xi''}^2\left|\xi''\right|^2 + \lambda_{\xi''} \left|\xi''\right|^4
\end{equation}

In the absence of cross-couplings, we seek minima that preserve the residual symmetries $\mathbb{Z}_2$ in the neutrino sector and $\mathbb{Z}_3$ in the charged-lepton sector. Working in the Altarelli-Feruglio (AF) basis, the minimization of the isolated potentials gives the standard alignments
$\langle \phi \rangle = v_\phi\begin{pmatrix}
    1 & 0 & 0
\end{pmatrix}^T$, $\langle \chi \rangle = \frac{v_\chi}{\sqrt{3}}\begin{pmatrix}
    1 & 1 & 1
\end{pmatrix}^T$, $\langle \eta \rangle = v_\eta$ and $\langle \xi'' \rangle = v_{\xi''}$. 
The VEV scales $v_\phi$ and $v_\chi$
 are determined strictly by the mass parameters and quartic couplings of their respective potentials, as derived in Appendix C below.\\

The renormalizable, $A_4 \otimes Z_4 \otimes GCP$ respecting cross-coupling between $\phi$ and $\chi$ is
\begin{equation}
    V(\phi\chi)=\frac{\epsilon_1}{2}(\phi \phi)_1(\chi \chi)_1 + \frac{\epsilon_2}{4}(\phi \phi)_{1''}(\chi \chi)_{1'}+\frac{\epsilon_2}{4}(\phi \phi)_{1'}(\chi \chi)_{1''}+\frac{\epsilon_3}{2}[(\phi \phi)_{3s} (\chi \chi)_{3s}]_1 + \frac{\epsilon_4}{3}[(\chi \chi)_{3s} \phi]_1
\end{equation}
with $\epsilon_1$, $\epsilon_2$, $\epsilon_3$, $\epsilon_4$ $\in$ $\mathbb{R}$ under GCP.

The renormalizable coupling between $\chi$ and $\xi''$ in our model is the main source connecting the non-zero $\theta_{13}$ and the bias term between $Z_2$ vacua, which collapses the domain wall and generates gravitational waves.
\begin{equation}
    V(\chi \xi'')=\kappa (\xi'')^2(\chi \chi)_{1''} + \kappa^* (\xi''^*)^2(\chi \chi)_{1'} 
\end{equation}
Here, $\kappa=\kappa_I +i\kappa_R$ $\in$ $\mathbb{C}$. This term is $A_4 \otimes Z_4$ invariant. Unlike $V_{cross}$, however, $V(\xi''\chi)$ is invariant under GCP only when $\kappa$ is real. For $\kappa_I \neq 0$, it constitutes explicit, dimension-four hard breaking of GCP. This is the sole new source of CP violation introduced by our model and, as shown in Appendix E, it simultaneously controls the reactor mixing angle $\theta_{13}$ and the amplitude of the domain-wall gravitational-wave bias. 

\section*{Appendix C}
Defining $g_1=f_1+f_2$, $g_2=\frac{3}{2}(f_3-f_2)$, $g_3=\sqrt{3}f_4$ and evaluating $V(\phi)$ in the AF basis $Z_3$ preserving representation $\langle \phi_- \rangle_\mp = \pm\sqrt{3} v_{\phi_\mp}\begin{pmatrix}
    1 & 0 & 0
\end{pmatrix}^T$ gives (using equations \eqref{eq:53})
\begin{equation}
    V|_{v\pm}=\frac{3\mu_\phi^2}{2}v_{\phi_\mp}^2 + \frac{3(3g_1+2g_2)}{4}v_{\phi_\mp}^4 \pm g_3 v_{\phi_\mp}^3
\end{equation}
Minimizing this gives,
\begin{equation}
    v_{\phi_\pm} =\frac{\mu_\phi}{\sqrt{3g_1+2g_2}}(\sqrt{1+a^2} \pm a)    
\end{equation}
Here, $a=\frac{g_3}{2\mu_\phi(\sqrt{3g_1+2g_2})}$ and the corresponding minimum energy 
\begin{equation}
    V_{v_{\phi_\mp}}|_{min}=-\frac{3(3g_1+2g_2)}{4}v_{\phi_\mp}^4 \mp \frac{g_3}{2}v_{\phi_\mp}^3
\end{equation}
Considering $v_{\phi_\pm}=v_0 u_\pm$, where $v_0=\frac{\mu_\phi}{\sqrt{3g_1+2g_2}}$, $u_\pm=\sqrt{1+a^2}\pm a$ and $u_+ u_-=1$, the bias between the two $Z_3$ preserving classes is
\begin{equation}
\begin{aligned}
     \Delta V_{Z_3} &=\frac{-4a\mu_\phi^4(1+a^2)^{3/2}}{(3g_1+2g_2)}\\
     \epsilon_b^{Z_3} &=\frac{4a(3g_1+2g_2)(1+a^2)^{3/2}}{(\sqrt{1+a^2}-a)^4}
\end{aligned}
\end{equation}
for small $a$, $\epsilon_b^{Z_3} \approx 4ag_1 \propto f_4$ and vanishes when $f_4=0$.

\section*{Appendix D}
\subsection*{Invariance of $\epsilon_\phi, \epsilon_\chi$ under $V(\chi \xi'')$}
Expanding about $\langle \phi \rangle = v_\phi \begin{pmatrix}
    1 & \epsilon_\phi & \epsilon_\phi
\end{pmatrix}^T$ and $\langle \chi \rangle = v_\phi \begin{pmatrix}
    1-2\epsilon_\chi,& 1+\epsilon_\chi,& 1+\epsilon_\chi
\end{pmatrix}^T$
and minimizing $V_{cross}$ gives $\epsilon_\phi$ and $\epsilon_\chi$ of the form identical to \cite{chen2026gravitational}, as shown below
\begin{equation}
\label{eq:63}
    \epsilon_\phi=\frac{3v_\chi^2 \epsilon_2}{(2\sqrt{3}f_4-6(f_2 + f_3 +2(f_1+a(a^2+\sqrt{1+a^2})f_1+a(a+\sqrt{1+a^2})f_3))v_{\phi_-})v_{\phi_-}}
\end{equation}
and
\begin{equation}
    \epsilon_\chi=\frac{v_{\phi-}(3\sqrt{3}v_{\phi-}\epsilon_3 +2\epsilon_4)}{3(4\tilde{f_1}+\tilde{f_2}+3\tilde{f_3})v_\chi^2}
\end{equation}
Since, $V(\chi \xi'')$ does not contain $\phi$ field, $\frac{\partial V(\chi \xi'')}{\partial \phi_i}=0$ identically, leaving equation ~\eqref{eq:63} unchanged.
Expanding ($\chi \chi)_{1''}$ about the perturbed $\chi$ vacuum,
\begin{equation}
    (\chi \chi)_{1''}=\frac{v_\chi^2}{3}[(1+\epsilon_\chi)^2+2(1-2\epsilon_\chi)(1+\epsilon_\chi)]=v_\chi^2
\end{equation}
independent of $\epsilon_\chi$ at 1st order, so $\frac{\partial V(\chi \xi'')}{\partial \epsilon_\chi}=0$. Thus $V(\chi \xi'')$ contribute no source for $\epsilon_\chi$.
\end{widetext}

\section*{Appendix E}
\subsection*{$Z_2$ Domain-Wall Bias from $V_{\chi \xi''}$ and Derivation of the $\theta_{13}$-GW relation}

At the three $Z_2$ preserving vacua $\langle \chi \rangle_j= \frac{v_\chi}{\sqrt{3}} (1, z_j, z_j^2)^T$, where $z_j=e^{2\pi i(j-1)/3}$, the tensor product of this triplet $\chi$ gives (for all $j$)
\begin{equation}
\begin{aligned}
    (\chi \chi)_1 &=v_\chi^2,\quad (\chi \chi)_{1'} =v_\chi^2 z_j\\
    (\chi \chi)_{1''} &=v_\chi^2 z_j^2
,\quad (\chi \chi)_{3_S} =0 \, .
\end{aligned}
\end{equation}
The vanishing of $(\chi \chi)_{3_S}$ in every sector implies the $\epsilon_4$ term of $V_{\rm cross}$ does not bias the $Z_2$ walls at leading order; but the sector-dependence of $(\chi \chi)_{1''}$ allows $V_{\chi\xi''}$ to do so.

Considering $\langle \xi'' \rangle =v_{\xi''}e^{i\delta}$ and substituting in Eq.~(58), we obtain
\begin{equation}
   V_{\xi'' \chi}(\delta)= 2|\kappa|v_{\xi''}^2 v_{\chi}^2 \cos{\left(2\delta + \phi_\kappa + \frac{4 \pi(j-1)}{3}\right)} ,
\end{equation}
where $\phi_\kappa=\arg(\kappa)$. Minimizing over $\delta$,
\begin{equation}
    \frac{\partial V_{\xi'' \chi}(\delta)}{\partial \delta}=-4|\kappa|v_{\xi''}^2 v_{\chi}^2 \sin{(2\delta + \phi_\kappa)}=0 \, ,
\end{equation}
gives the stable phase
\begin{equation}
        2\delta+\phi_\kappa =\pi \implies 
        \delta=\frac{\pi-\phi_\kappa}{2} \, .
        \label{eq:delta_exact}
\end{equation}
Writing $\kappa=\kappa_R+i\kappa_I$, Eq.~\eqref{eq:delta_exact} admits two GCP-conserving stationary points, corresponding to the two solutions of $\kappa_I=0$:

 \textbf{Branch A} ($\kappa_R<0$): $\phi_\kappa=\pi \implies \delta=0$, so $\langle \xi'' \rangle=v_{\xi''} \in \mathbb{R}$, GCP is preserved and TBM is exact. Linearizing for $|\kappa_I|\ll|\kappa_R|$ about this branch gives $\phi_\kappa \approx \pi-\kappa_I/|\kappa_R|$, and hence
\begin{equation}
    \delta \approx \frac{\kappa_I}{2|\kappa_R|} \qquad \text{(valid near } \delta=0 \text{)}.
    \label{eq:branchA}
\end{equation}

\textbf{Branch B} ($\kappa_R>0$): $\phi_\kappa=0 \implies \delta=\pi/2$, an equally valid GCP-conserving stationary point. Linearizing for $|\kappa_I|\ll|\kappa_R|$ about this branch gives $\phi_\kappa \approx \kappa_I/|\kappa_R|$, and hence
\begin{equation}
    \delta \approx \frac{\pi}{2}-\frac{\kappa_I}{2\kappa_R} \qquad \text{(valid near } \delta=\pi/2 \text{)}.
    \label{eq:branchB}
\end{equation}

Our global best fit yields $\delta = 93.53^\circ$, which lies close to $\pi/2$ rather than $0$; the fit therefore selects \textbf{Branch B}. We can see that , Eq.~\eqref{eq:branchB} gives $\kappa_I/\kappa_R \approx -0.123$, in agreement to within $0.4\%$ with the exact relation $\kappa_I/\kappa_R=\tan\phi_\kappa$ evaluated at $\phi_\kappa=\pi-2\delta=-7.06^\circ$. All subsequent formulas are therefore expanded about Branch B.

Substituting the stable minimum, Eq.~\eqref{eq:delta_exact}, into the vacuum energy expression, we obtain
\begin{equation}
    \langle V_{\xi'' \chi} \rangle_{j=1} = -2|\kappa|v_{\xi''}^2 v_{\chi}^2, \quad  \langle V_{\xi'' \chi} \rangle_{j={2,3}} =|\kappa|v_{\xi''}^2 v_{\chi}^2 \, .
\end{equation}
This gives a new bias term between the $Z_2$ vacua,
\begin{equation}
    \epsilon_b^{\xi''}=\frac{3|\kappa|v_{\xi''}^2}{v_{\chi}^2} \, .
    \label{eq:epsb_exact}
\end{equation}
Now, using the relation of Eq.~(33) together with $c_1=\frac{y_{\xi''} v_{\xi''}}{2}$ and $b_1=\frac{y_{N} v_{\chi}}{2\sqrt{3}}$, Eq.~\eqref{eq:epsb_exact} may be rewritten as
\begin{equation}
    \epsilon_b^{\xi''}=\frac{216 |\kappa|b_1^2}{y_{\xi''}^2 v_\chi^2 \sin^2{\delta}}\sin^2{\theta_{13}} \, .
    \label{eq:epsb_delta}
\end{equation}

To obtain a closed-form relation between $\epsilon_b^{\xi''}$ and $\sin^2\theta_{13}$ alone, we evaluate $\sin^2\delta$ using the correct branch identified above. Writing $\delta = \pi/2 - x$ with $x \equiv \kappa_I/(2\kappa_R)$ from Eq.~\eqref{eq:branchB}, and expanding for small $x$,
\begin{equation}
\begin{aligned}
    \sin^2\delta &= \sin^2\left(\frac{\pi}{2}-x\right) = \cos^2x = 1-x^2+\mathcal{O}(x^4)\\ &= 1-\left(\frac{\kappa_I}{2\kappa_R}\right)^2+\mathcal{O}\!\left(\frac{\kappa_I}{\kappa_R}\right)^4 \, .
\end{aligned}
\end{equation}
Since $|\kappa_I/\kappa_R|\ll1$ throughout the viable parameter space, $\sin^2\delta \approx 1$ to leading order, and Eq.~\eqref{eq:epsb_delta} simplifies to
\begin{equation}
    \epsilon_b^{\xi''} \approx \mathcal{K}\sin^2{\theta_{13}} \, , \qquad
    \mathcal{K}=\frac{216\,|\kappa|\,b_1^2}{y_{\xi''}^2 v_\chi^2} = \frac{18\,|\kappa|\,y_N^2}{y_{\xi''}^2} \, ,
    \label{eq:master}
\end{equation}
Corrections to Eq.~\eqref{eq:master} are suppressed by $(\kappa_I/\kappa_R)^2 \sim \mathcal{O}(10^{-2})$ at our benchmark point.

We see that the coefficient 
$\mathcal{K}$ in Eq.~\eqref{eq:master} remains finite in the limit $\kappa_I\to0$ on 
Branch B. It is important to note that Eq.~\eqref{eq:delta_exact} admits two physically 
distinct, exactly GCP-conserving solutions at $\kappa_I=0$: Branch A ($\delta=0$), at which 
$U_{e3}=0$ and TBM is exact, and Branch B ($\delta=\pi/2$), at which 
$\sin\theta_{13}$ remains fixed by the mass-sector parameter $c$ while 
$U_{e3}$ is purely imaginary, protecting $\delta_{CP}=270^\circ$ 
exactly. Both points are genuine fixed points of the GCP symmetry, since a field 
carrying nontrivial $Z_4$ charge, in our case $\xi''$, generically admits both trivial 
($\delta_{CP}\in\{0,\pi\}$) and maximal ($\delta_{CP}\in\{\pi/2,3\pi/2\}$) CP-conserving 
vacua. Our global fit selects Branch B, so GCP restoration ($\kappa_I\to0$) does not send 
$\theta_{13}\to0$; instead it sends the deviation of $\delta_{CP}$ away from its 
GCP-protected value $270^\circ$ to zero, explaining the model's prediction of near-maximal 
Dirac CP violation as a symmetry-protected outcome rather than an accidental feature of the 
fit. On this branch, $\theta_{13}$ is set predominantly by $c$, while $\kappa$ controls both 
the small deviation of $\delta_{CP}$ from $270^\circ$ and, through $|\kappa|$, the amplitude 
of the $\mathbb{Z}_2$ domain-wall bias.

\appendix

\nocite{*}

\bibliography{apssamp}

\end{document}